\documentclass[conference]{IEEEtran}
\IEEEoverridecommandlockouts
\usepackage{graphicx} 
\usepackage{cite}
\usepackage[colorlinks=true, citecolor=blue, linkcolor=blue, urlcolor=blue]{hyperref}
\usepackage{amsmath,amssymb,amsfonts}
\usepackage{algorithmic}
\usepackage{graphicx}
\usepackage{textcomp}
\usepackage{xcolor}
\def\BibTeX{{\rm B\kern-.05em{\sc i\kern-.025em b}\kern-.08em
    T\kern-.1667em\lower.7ex\hbox{E}\kern-.125emX}}
\begin{document}

\title{Time Domain Near Memory Computing Engine\\}

\author{
\begin{tabular}{cc}
\begin{minipage}{0.46\textwidth}
\centering
Sarthak Antal\\
\textit{Ph.D. Student}\\
\textit{Elmore Family School of ECE}\\
\textit{Purdue University}\\
West Lafayette, USA\\
santal@purdue.edu
\end{minipage}
&
\begin{minipage}{0.46\textwidth}
\centering
Steve Enosh\\
\textit{M.S. Student}\\
\textit{Elmore Family School of ECE}\\
\textit{Purdue University}\\
West Lafayette, USA\\
manda4@purdue.edu
\end{minipage}
\end{tabular}
}
\maketitle

\begin{abstract}
The increasing computational demand of AI workloads has intensified the need for energy-efficient in-memory and near-memory computing architectures, particularly because data movement often consumes significantly more energy than the computation itself. While fully digital architectures provide robust scalability and support higher-resolution computation, analog in-memory computing has demonstrated improved energy efficiency for low-precision workloads. However, its reliance on peripheral DACs and ADCs introduces additional power, area, and design overhead. To address these challenges, this work presents a time-domain near-memory computing architecture for low-precision multiply-and-accumulate (MAC) operations. In the proposed approach, digital weight bits stored in SRAM are converted using a current-steering DAC, while the digital input vector is encoded by an N-pulse generator. This enables multiplication to be performed in the time domain while maintaining a digital-friendly interface. Two accumulation schemes, a delay-cell-based architecture and a counter-based architecture, are investigated and compared in terms of design trade-offs, linearity, scalability, and power efficiency. To improve technology portability, the N-pulse generator and counters are implemented using RTL synthesis, while the current-steering DAC remains in the analog domain. A 4 × 4 MAC prototype is implemented in a 1 V supply, achieving an operating frequency of 40 MHz, power consumption of 42uW , and energy efficiency of 7.62 TOPS/W.

\end{abstract}


\section{Introduction}

With the rapid growth of AI workloads, the demand for energy-efficient in-memory and near-memory computing has increased significantly, as the energy cost of data movement can be orders of magnitude higher than that of computation. In the target architecture, the weights are stored in SRAM, while the input vectors are provided in digital form. Analog computing architectures suffer an overhead cost of additional peripheral circuits, such as DACs and ADCs, since the input and output interfaces of the computing engine must remain digital. Fully digital architectures offer an attractive solution for higher-resolution computation, typically beyond 8 bits. However, recent studies have shown that for applications requiring sub-8-bit precision, analog in-memory computing can provide superior energy efficiency. \cite{b1}. The energy efficiency of voltage-domain data converters degrades rapidly with increasing resolution. For an ideal ADC, the signal-to-noise ratio (SNR) is related to the effective number of bits (ENOB) as
\begin{equation}
    \mathrm{SNR} = 6.02 \times \mathrm{ENOB} + 1.76 \ \mathrm{dB}.
\end{equation}

Therefore, improving the resolution by 1 bit requires approximately a 6 dB improvement in SNR. Since SNR is proportional to the ratio of signal power to noise power, a 6 dB SNR improvement corresponds to a 4$\times$ reduction in noise power. In voltage-domain circuits, the dominant noise contribution is often thermal noise, which is proportional to $kT/C$. Thus, reducing the noise power by 4$\times$ requires increasing the sampling capacitance by approximately 4$\times$.

Since the dynamic power required to drive capacitive loads scales linearly with capacitance, a 4$\times$ increase in capacitance results in approximately a 4$\times$ increase in power consumption. Consequently, achieving each additional bit of resolution in a voltage-domain ADC typically requires an approximately 4$\times$ increase in power. This exponential dependence highlights the fundamental power-scaling limitation of voltage-domain data converters at higher resolutions.

With continued technology scaling, the reduced voltage headroom limits the achievable SNR of ADCs and DACs, as the available signal power decreases. One approach to address this challenge is heterogeneous integration, where different technology nodes can be optimized for digital and analog functions independently. However, this approach introduces additional overheads, including high-speed I/O interfaces, increased power consumption, integration complexity, and fabrication cost.

Alternatively, recent research has explored hybrid voltage–time-domain \cite{b2} and fully time-domain architectures \cite{b3}. These approaches exploit the fact that the device transition frequency (ft) generally improves with technology scaling, enabling faster switching, improved timing resolution, and enhanced jitter tolerance. As a result, time-domain circuits, particularly time-to-digital converters (TDCs), have emerged as a promising solution for achieving higher resolution in scaled technologies.

In this work, we propose a time-domain near-memory computing architecture in which the digital weight bits fetched from SRAM are applied to a current-steering DAC, while the input vector is encoded using an N-pulse generator. This enables multiplication to be performed in the time domain with improved power efficiency. For the accumulation operation, two architectures are investigated: a delay-cell-based approach and a non-traditional counter-based adding approach. Their design trade-offs are analyzed, and the results of the superior architecture are presented. 

The paper is organized as follows: Section II presents the system-level design of the two proposed macro architectures. Section III describes the timing diagram of the Macros. Section IV describes the individual circuit blocks and their corresponding simulation results. Section V presents compares the two architectures.  Section VI is the top-level simulation results and concludes the paper in Section VII by comparing the proposed design with prior state-of-the-art works and outlining potential directions for future work.

\section{System Level Design}

\subsection{Architecture I: Delay-Cell-Based Time-Domain MAC}

The first architecture  as shown in Fig1 performs the MAC operation in two phases: a multiplication phase and an accumulation phase. In the multiplication phase, a $4 \times 4$ multiplication is performed between the digital input vector and the digital weight values. Since the inputs are available in the digital domain, the input vector is first converted into the time domain using an $N$-pulse generator. The $N$-pulse generator is a digital block that produces a number of pulses proportional to the input code. For example, an input code of $0000$ produces no pulse activity, while an input code of $1111$ produces 16 pulses. Thus, the 4-bit digital input is encoded as a pulse-count-based time-domain signal.

The $N$-pulse generator is implemented using RTL synthesis and designed with timing constraints that support a maximum clock frequency of 100 MHz. Therefore, the minimum pulse width is determined by the clock period corresponding to this operating frequency. In parallel, the digital weight bits are applied to a 4-bit current-steering DAC, which generates an output current proportional to the digital weight code.

The multiplication operation is realized by using the DAC output current and the pulse duration generated by the $N$-pulse generator to charge a capacitor. The resulting voltage across the capacitor is given by
\begin{equation}
    V_{\mathrm{MAC}} = \frac{I_{\mathrm{DAC}} \cdot T_{\mathrm{pulse}}}{C},
\end{equation}
where $I_{\mathrm{DAC}}$ is the current generated by the current-steering DAC, $T_{\mathrm{pulse}}$ is the effective pulse duration generated by the $N$-pulse generator, and $C$ is the integration capacitance. The value of $C$ is chosen such that the thermal noise contribution remains below the quantization noise corresponding to the target 4-bit resolution. The thermal noise associated with the capacitor is given by
\begin{equation}
    \overline{v_n^2} = \frac{kT}{C},
\end{equation}
where $k$ is Boltzmann's constant and $T$ is the absolute temperature.

The accumulation operation is performed using a delay-cell-based architecture. The voltage generated from each input--weight multiplication controls a PMOS device that current-starves a chain of eight inverters. The output of each delay cell is then applied to the input of the next delay stage, whose delay is controlled by the subsequent input--weight product. In this manner, the delay contributions from multiple multiplication results are accumulated in the time domain, thereby realizing the accumulation operation.

The final digital output is obtained using a control logic block and a counter, which measure the delay between the input rising edge and the output rising edge of the accumulated delay chain. This measured delay represents the accumulated MAC result in the digital domain.

However, a key limitation of this architecture is its degraded linearity as the matrix dimension increases. Since the accumulation is performed by cascading multiple voltage-controlled delay cells, the nonlinear delay characteristics of each stage accumulate across the chain. As a result, the overall transfer characteristic becomes increasingly distorted for larger MAC arrays. The detailed circuit implementation and simulation results of this architecture are discussed in the following sections.

\begin{figure}[htbp]
    \centering
    \includegraphics[scale=0.28]{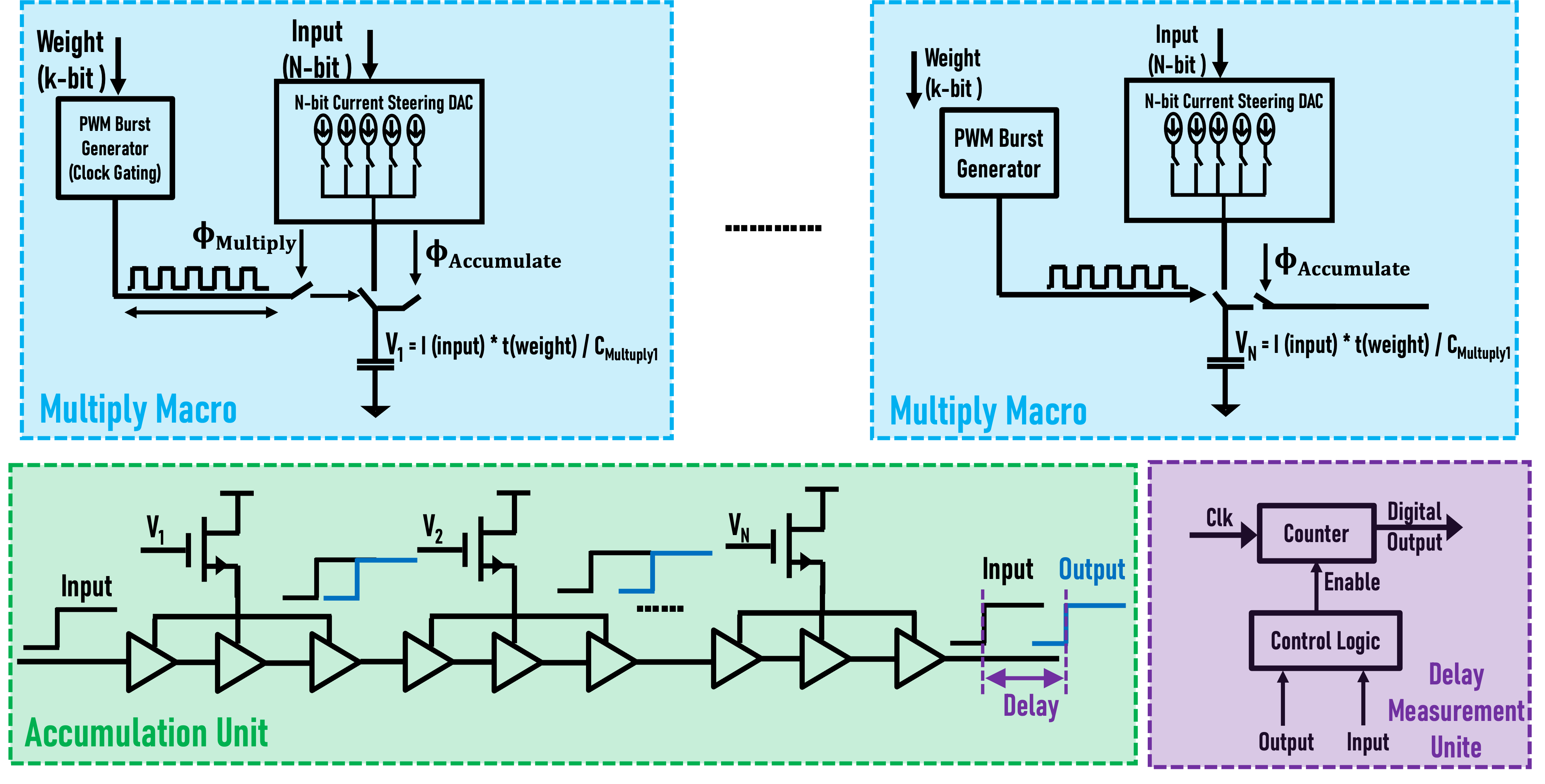} 
    \caption{Delay Cell based Macro}
    \label{fig:1}
\end{figure}

\subsection{Architecture II: Counter-Based Time-Domain MAC}

The second architecture as shown in Fig 2 uses the same multiplication scheme as Architecture I; however, the accumulation operation is implemented using a counter-based approach instead of a cascaded delay-line structure. As described previously, each multiplication macro converts the digital input and weight values into a voltage-domain product. This multiplication output voltage is applied to a PMOS current-starving device, which controls the delay of an inverter-based delay cell.

Unlike Architecture I, the delay cells are not cascaded for direct time-domain accumulation. Instead, each delay cell is evaluated independently, and its corresponding delay is converted into a digital count. A control logic block sequentially selects each multiplication macro and enables the corresponding delay cell. For each selected cell, the counter measures the time interval between the input transition and the delayed output transition. After the delay of one cell is measured, the control logic disables the current cell and applies the input transition to the next delay cell.

In the present implementation, the input transitions to the individual delay cells are generated using ideal voltage sources from the analog library for functional verification. The measured delay count from each cell is accumulated digitally in the counter, thereby realizing the accumulation operation. If $t_{d,i}$ represents the delay generated by the $i$th multiplication cell, the ideal accumulated delay corresponding to an $N$-cell MAC operation can be expressed as
\begin{equation}
    t_{d,\mathrm{acc}} = \sum_{i=1}^{N} t_{d,i}.
\end{equation}

In the counter-based architecture, each delay is quantized by the counter clock period, $T_{\mathrm{clk}}$. Therefore, the measured delay of the $i$th cell can be written as
\begin{equation}
    \hat{t}_{d,i} = T_{\mathrm{clk}} 
    \left\lfloor \frac{t_{d,i}}{T_{\mathrm{clk}}} \right\rfloor,
\end{equation}
where $\hat{t}_{d,i}$ is the quantized delay measured by the counter. The corresponding quantization error is given by
\begin{equation}
    e_{q,i} = t_{d,i} - \hat{t}_{d,i},
\end{equation}
with the error bounded as
\begin{equation}
    0 \leq e_{q,i} < T_{\mathrm{clk}}.
\end{equation}

Assuming the quantization error is uniformly distributed over one clock period, the timing-domain quantization-noise variance is
\begin{equation}
    \sigma_{q,t}^{2} = \frac{T_{\mathrm{clk}}^{2}}{12}.
\end{equation}

For an accumulated MAC operation consisting of $N$ independently measured delay cells, the accumulated timing-noise variance becomes

\begin{figure}[htbp]
    \centering
    \includegraphics[scale=0.28]{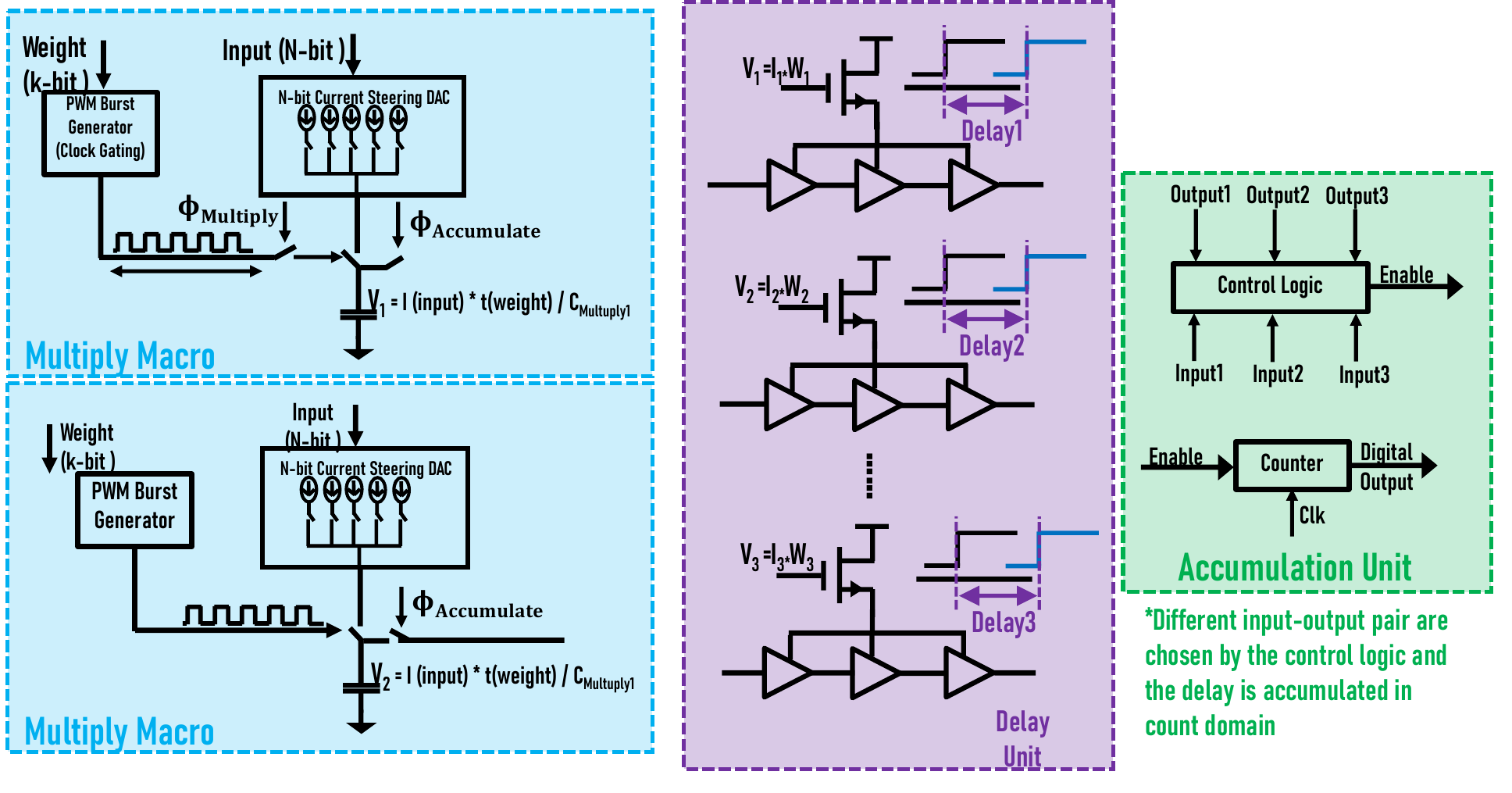} 
    \caption{Delay Cell based Macro}
    \label{fig:2}
\end{figure}

\begin{equation}
    \sigma_{q,t,\mathrm{acc}}^{2}
    = \sum_{i=1}^{N} \sigma_{q,t,i}^{2}
    = N \frac{T_{\mathrm{clk}}^{2}}{12}.
\end{equation}

The final accumulated counter output can therefore be represented as
\begin{equation}
    D_{\mathrm{out}} = \sum_{i=1}^{N}
    \left\lfloor \frac{t_{d,i}}{T_{\mathrm{clk}}} \right\rfloor .
\end{equation}

Thus, the accuracy of the counter-based accumulation is primarily limited by the counter clock resolution. Increasing the counter clock frequency reduces $T_{\mathrm{clk}}$, thereby reducing the timing-domain quantization noise and improving the effective resolution of the accumulated MAC output.

This approach avoids the direct cascading of multiple nonlinear delay cells and prevents the delay error of one stage from propagating into subsequent stages. As a result, the counter-based architecture provides improved scalability and better linearity compared with the delay-line-based accumulation scheme, particularly as the matrix dimension increases. The final MAC output is represented by the accumulated counter value, which corresponds to the sum of the individual input--weight products. The detailed circuit implementation, control sequence, and simulation results of this architecture are discussed in the following sections.

\section{Timing Diagram}
This section describes the operational principles of the proposed time-domain compute macros, which performs multiply accumulate (MAC) operations in three distinct phases.

\paragraph{Multiplication Phase}
In the first phase, input activations and weights are encoded into the analog domain. Multiplication is inherently realized within each compute macro through the interaction of these analog representations, enabling parallel and energy-efficient computation.

\paragraph{Accumulation Phase}
During the accumulation phase, the partial products generated in the previous stage are aggregated. The specific accumulation mechanism is architecture-dependent and varies with the chosen macro design, impacting both precision and energy efficiency.

\paragraph{Reset Phase}
In the final phase, the capacitor nodes are reset by discharging the stored voltage to zero. This ensures proper initialization and prevents residual charge from affecting subsequent computation cycles.

\begin{figure*}[t]
    \centering
    \includegraphics[width=0.9\textwidth]{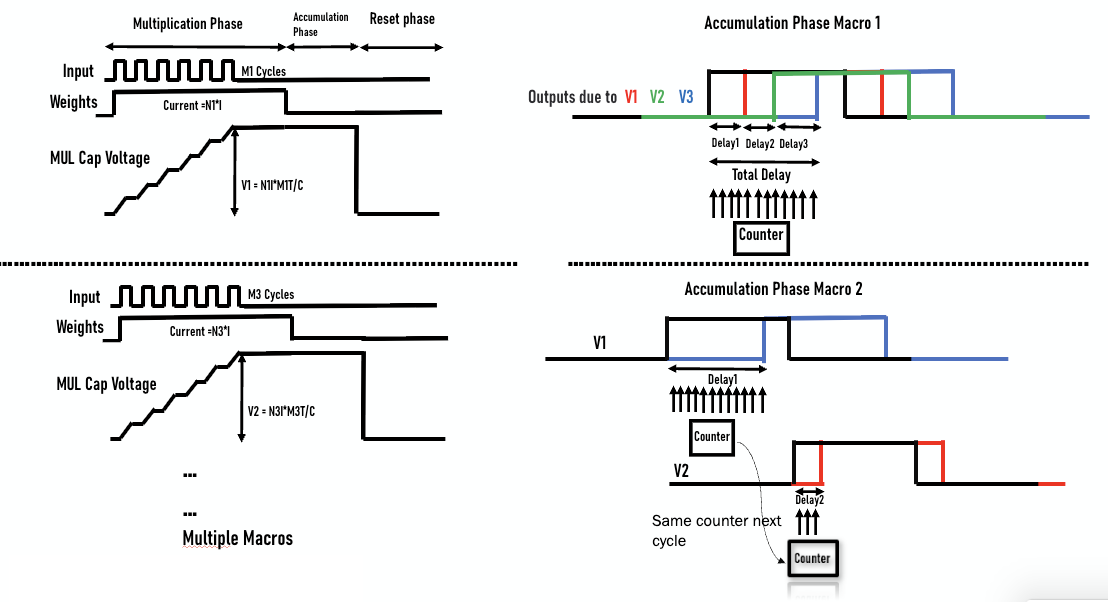}
    \caption{Timing diagram of the macros.}
    \label{fig:3}
\end{figure*}

\section{Implementation of Building blocks}

\subsection{Current Steering DAC}

The 4-bit current-steering DAC (CSDAC) is designed to convert digital input into analog currents, which are then integrated onto capacitors to generate a corresponding voltage for accumulation. The CSDAC is implemented using current mirrors, differential switching pairs, and accumulation capacitors. To mitigate nonlinearity arising from finite output impedance, the current mirror is cascoded, thereby significantly increasing the effective output resistance of the current sources. The unit current (LSB) is designed to be 11.5 nA, resulting in a full-scale current of 166 nA for 4-bit resolution. During operation, charge is dumped onto the capacitors, causing the voltage at the bottom node of the switch transistors to rise. To ensure sufficient voltage headroom and maintain proper operation of the switches, PMOS transistors are used, as they provide adequate VSD under these conditions, helping to preserve saturation region. Furthermore, to limit nonlinearity, the maximum allowable voltage swing across the accumulation capacitors is constrained to approximately 300 mV. This ensures that the current sources operate in saturation region, thereby maintaining good linearity for the multiplications.

\begin{figure}[htbp]
    \centering
    \includegraphics[scale=0.18]{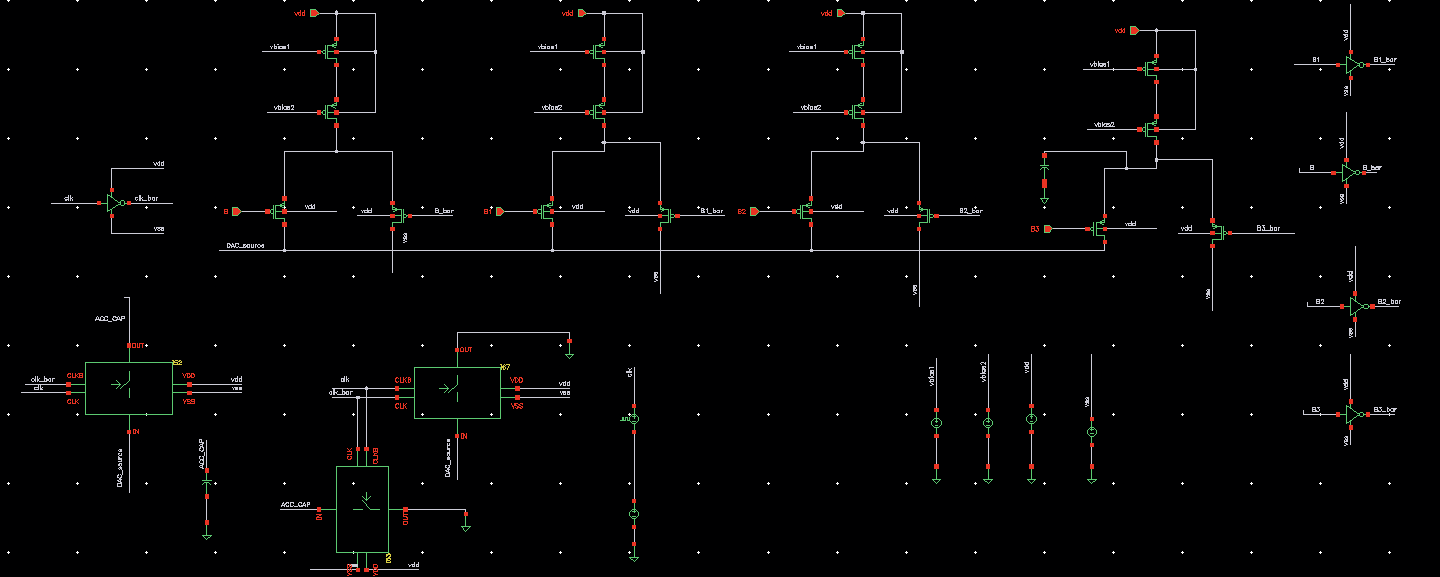} 
    \caption{Schematic of Current Steering DAC}
    \label{fig:4}
\end{figure}
\begin{figure}[htbp]
    \centering
    \includegraphics[scale=0.25]{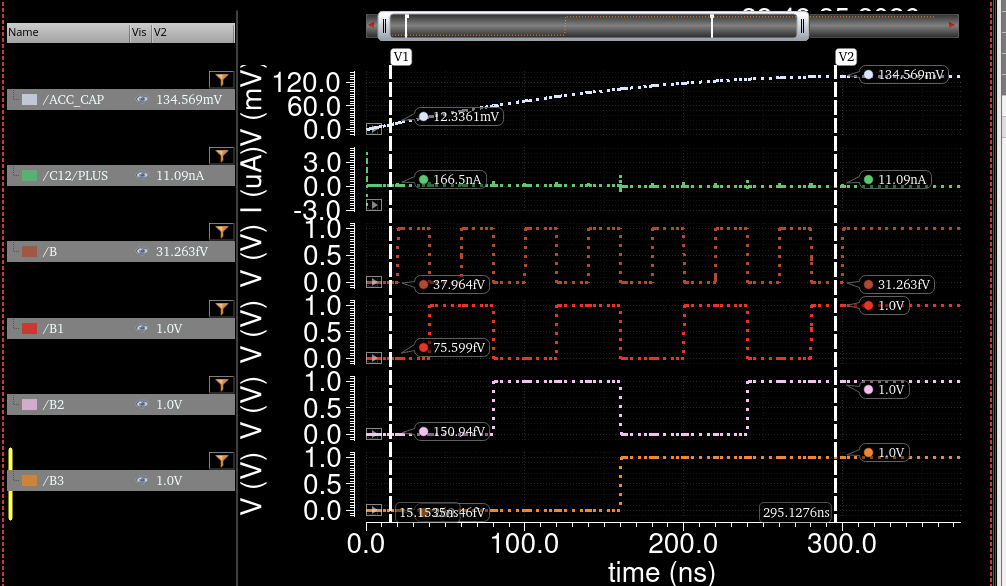} 
    \caption{Output of Current Steering DAC sweeping codes from 1111 to 0000}
    \label{fig:5}
\end{figure}

\subsection{Accumulation capacitor}

The multiplication partial product is the voltage produced by the dumped current on the accumulation capacitor. The voltage on the capacitor should not exceed 300mV to maintain saturation of transistors. This puts a constraint on current of CSDAC and total ON time of PWM DAC pulses. The size of the capacitor also plays a huge role in determining the voltage.

\paragraph{}
The relationship between voltage, current, and integration time in the proposed architecture is given by
\begin{equation}
    V = \frac{I \cdot T}{C}.
\end{equation}

For the maximum input condition, the integrated voltage is constrained to a maximum allowable value of 300\,mV. Assuming a full-scale current of 165\,nA and a capacitance of 200\,fF, the maximum achievable integration (ON) time can be derived as
\begin{equation}
    T_{\text{max}} = \frac{V_{\text{max}} \cdot C}{I_{\text{max}}}
    = \frac{300\,\text{mV} \times 200\,\text{fF}}{165\,\text{nA}}
    \approx 363\,\text{ns}.
\end{equation}

Accordingly, the maximum unit pulse width is limited to 20 ns. This highlights an inherent trade-off between the capacitor size and the pulse width of the PWM DAC, which directly impacts the achievable dynamic range and timing resolution of the system.

\subsection{PWM DAC: N-Pulse Generator Control Logic}

The $N$-pulse generator is implemented using a counter-based digital control block. The purpose of this block is to generate a pulse train whose duration, or number of clock cycles, is proportional to the input digital code. The input code is stored in a 4-bit register, while a 4-bit counter increments on every clock cycle when the enable signal is asserted. The counter value is continuously compared with the stored input code. When the counter value matches the programmed register value, the output control signal is de-asserted, thereby terminating the pulse generation window.

The RTL implementation of the control block consists of a 4-bit counter, a 4-bit input register, and a comparator. The module takes a clock signal, an active-low reset signal, an enable signal, and a 4-bit digital input code as inputs. The corresponding counter value, registered input code, and match signal are provided as outputs. The match signal remains high during the pulse-generation interval and transitions low once the counter reaches the programmed input code.

During reset, the counter and register are initialized to zero. The match signal is initialized high so that the output pulse window starts from a known active state after reset. A valid comparison flag is also used to prevent an incorrect match condition immediately after reset, since both the counter and register are initially zero. This avoids a false $0 = 0$ comparison before the input code has been properly loaded.

After reset is released, the input code is sampled into the internal register on the rising edge of the clock. When the enable signal is high, the counter increments by one count on each clock cycle. The comparator checks whether the counter value is equal to the stored register value. Once the equality condition is satisfied, the match output is pulled low and remains low. Therefore, the high duration of the match signal represents the number of clock cycles corresponding to the digital input code.

The output pulse duration can be expressed as
\begin{equation}
    T_{\mathrm{pulse}} = N_{\mathrm{in}} T_{\mathrm{clk}},
\end{equation}
where $N_{\mathrm{in}}$ is the digital input code and $T_{\mathrm{clk}}$ is the clock period of the pulse generator. Equivalently, for a clock frequency $f_{\mathrm{clk}}$, the pulse duration is given by
\begin{equation}
    T_{\mathrm{pulse}} = \frac{N_{\mathrm{in}}}{f_{\mathrm{clk}}}.
\end{equation}

Thus, the digital input code is converted into a time-domain representation. A smaller input code produces a shorter pulse-generation window, while a larger input code produces a longer pulse-generation window. For a 4-bit input, the pulse generator can represent up to 16 discrete timing levels. For example, an input code of $0000$ corresponds to no pulse activity, while the maximum input code corresponds to the longest pulse duration. This time-domain encoded signal is then used in the multiplication macro, where it controls the integration time for charging the capacitor with the current generated by the current-steering DAC.

\begin{figure}[htbp]
    \centering
    \includegraphics[scale=0.25]{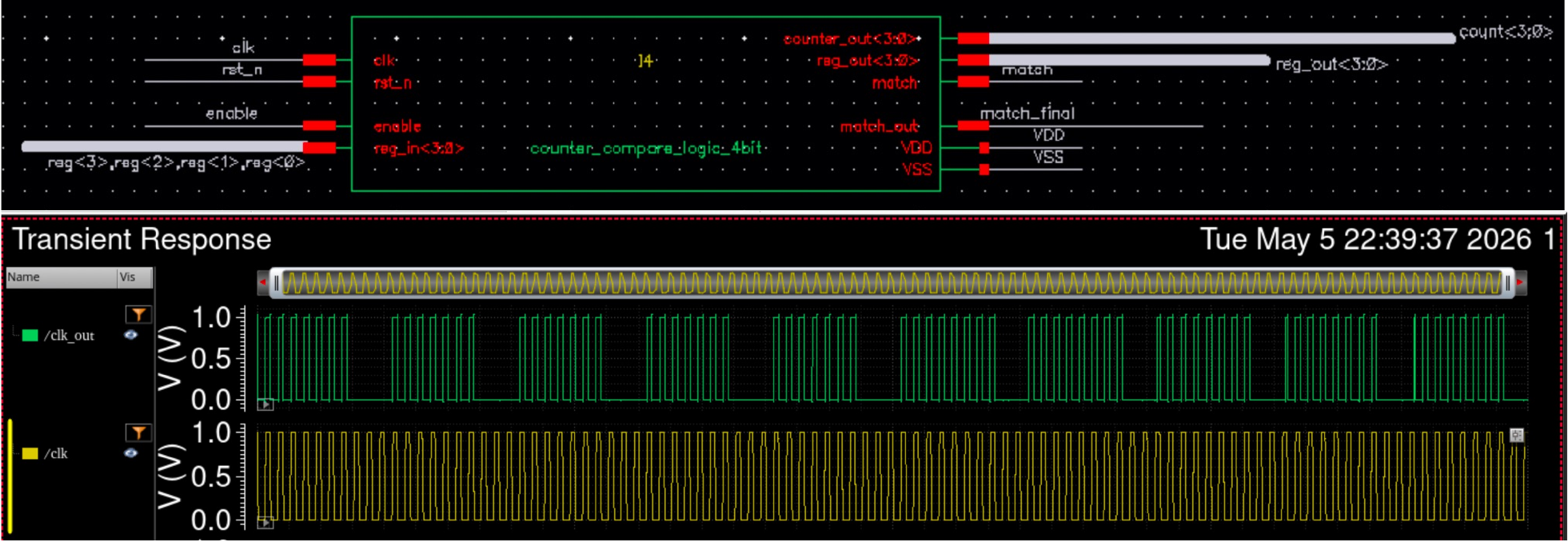} 
    \caption{Output of N Pulse Generator}
    \label{fig:6}
\end{figure}

Fig.~6 shows the synthesized implementation of the proposed $N$-pulse generator. The clock signal is applied to the synthesized digital block, while the 4-bit input code determines the number of output pulses generated. In the example shown, the input code is set to $0111$, corresponding to a decimal value of 7. As a result, the counter increments on each clock cycle and the comparator keeps the output active until the counter value matches the registered input code. Therefore, seven output pulses are generated, demonstrating the conversion of the digital input code into a pulse-count-based time-domain representation.

\subsection{Current-Starved Delay Cell}

The time-domain conversion in the proposed MAC architecture is performed using a current-starved delay cell. Each delay cell consists of a chain of eight cascaded inverters whose propagation delay is controlled by a PMOS current-starving device. The multiplication output voltage generated by the capacitor-charging operation is applied to the gate of the PMOS device, thereby modulating the available charging current of the inverter chain. As a result, the voltage-domain multiplication result is converted into a delay-domain representation.

For an inverter driving an effective load capacitance $C_{\mathrm{L}}$, the propagation delay can be approximated as
\begin{equation}
    t_{d} \approx \frac{C_{\mathrm{L}} V_{\mathrm{sw}}}{I_{\mathrm{starve}}},
\end{equation}
where $V_{\mathrm{sw}}$ is the effective output voltage swing and $I_{\mathrm{starve}}$ is the current provided through the PMOS current-starving device. Since the current-starving PMOS is controlled by the multiplication output voltage, the delay becomes a function of the MAC product voltage,
\begin{equation}
    t_{d} = f(V_{\mathrm{MAC}}).
\end{equation}

For an eight-inverter delay chain, the total delay can be expressed as
\begin{equation}
    t_{d,\mathrm{cell}} = \sum_{k=1}^{8} t_{d,k}
    \approx 8 \frac{C_{\mathrm{L}} V_{\mathrm{sw}}}{I_{\mathrm{starve}}(V_{\mathrm{MAC}})}.
\end{equation}

The linearity of this voltage-to-delay conversion is limited by the operating range of the PMOS current-starving device. In the designed delay cell, the PMOS exhibits an approximately linear current-control region only up to around 300 mV of control-voltage variation. Therefore, the capacitor-charging output from the multiplication stage is constrained to remain within this linear range.

This constraint directly limits the maximum supported multiplication range. In the proposed 4-bit by 4-bit multiplication, the maximum input and weight values correspond to 16 discrete levels each, resulting in a maximum product of
\begin{equation}
    P_{\mathrm{max}} = 16 \times 16 = 256.
\end{equation}
The multiplication capacitor and current-scaling factors are therefore selected such that the maximum product maps to a control voltage of approximately 256 mV, which remains below the 300 mV linearity limit of the PMOS current-starving device:
\begin{equation}
    V_{\mathrm{MAC,max}} \approx 256~\mathrm{mV} < 300~\mathrm{mV}.
\end{equation}

This scaling ensures that the delay cell operates within its approximately linear voltage-to-delay conversion region for the full 4-bit input and 4-bit weight range. However, extending the architecture to larger input precision or larger matrix dimensions would increase the required multiplication or accumulation range, forcing the PMOS control voltage beyond its linear region. This would introduce significant delay nonlinearity and degrade the accuracy of the time-domain MAC operation.

During operation, a rising edge is applied to the input of the delay cell. The controlled inverter chain delays this transition by an amount determined by $V_{\mathrm{MAC}}$. The delay between the input rising edge and the output rising edge is then measured by the time-to-digital conversion circuitry. Therefore, the delay cell acts as a voltage-to-time converter, translating the analog multiplication result into a time-domain quantity that can be digitized by the TDC or counter-based readout.

\begin{figure}[htbp]
    \centering
    \includegraphics[scale=0.25]{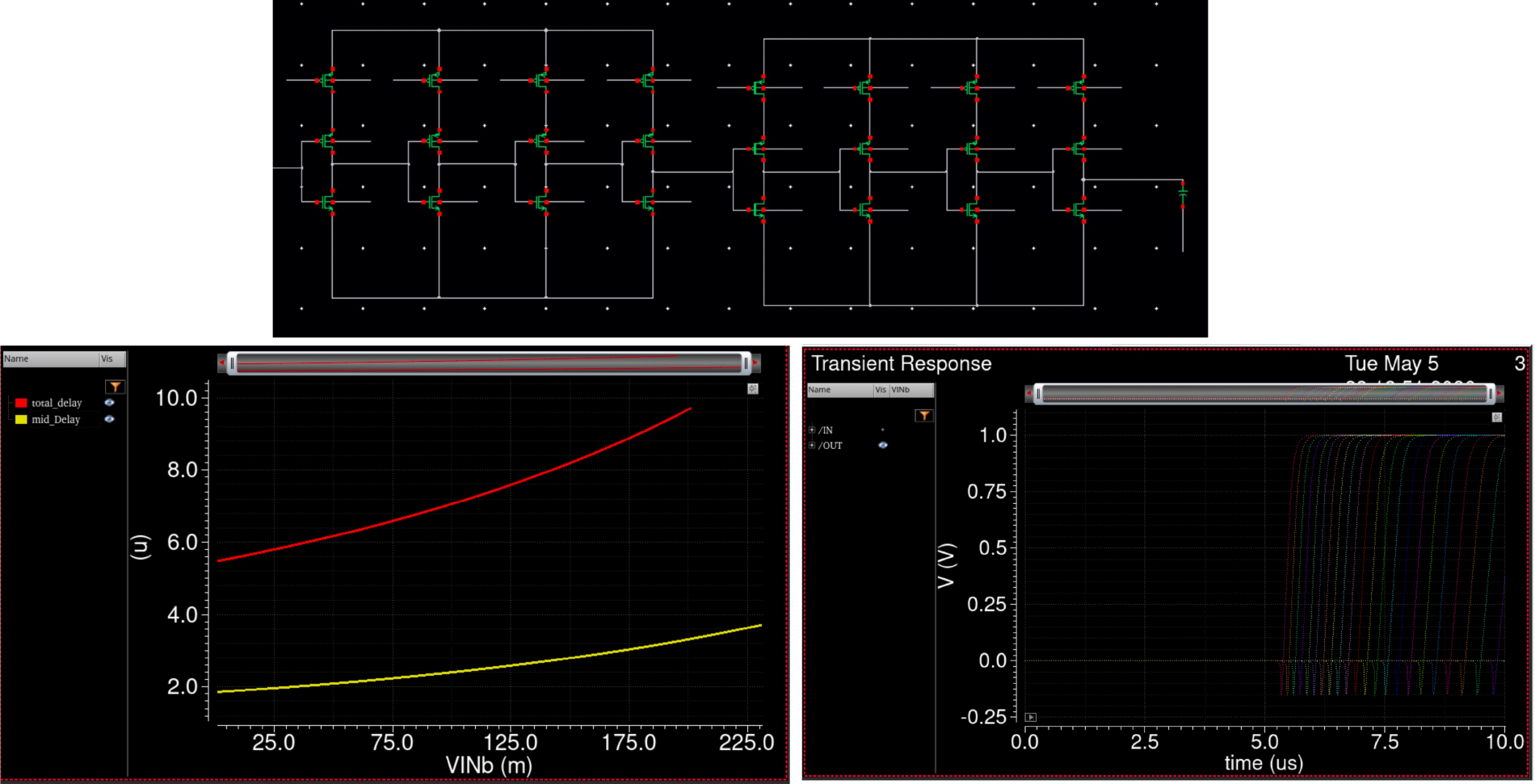} 
    \caption{Delay Cell Schematic and Results}
    \label{fig:7}
\end{figure} 
Fig.~8 illustrates the implementation and operation of the proposed current-starved delay cell. The top portion of the figure shows the schematic of a single delay cell, which consists of eight cascaded inverters current-starved by a PMOS device. The gate voltage of the PMOS is controlled by the multiplication output voltage, thereby modulating the delay of the inverter chain.

The bottom-right plot shows the delay variation measured at different nodes of the delay cell. Specifically, the delay is observed at an intermediate node as well as at the final output node of the inverter chain. As expected, the final output node exhibits a larger delay variation because the delay contribution from each inverter stage accumulates along the chain. This confirms that the delay cell converts the control-voltage variation at the PMOS gate into a measurable time-domain delay.

The remaining plot shows the delayed output waveforms for different PMOS control voltages. As the PMOS gate voltage changes, the current available to the inverter chain is modulated, resulting in different propagation delays. Therefore, the rising edge at the output shifts in time depending on the multiplication output voltage applied to the PMOS. This behavior demonstrates the voltage-to-time conversion mechanism used in the proposed time-domain MAC architecture.

\section{Comparison between the architectures}

\subsection{Comparison Between the Two Accumulation Architectures}

The two proposed accumulation architectures present a fundamental trade-off between throughput, linearity, circuit complexity, and power efficiency. In the cascaded delay-line architecture, the accumulation is performed directly in the time domain by propagating a single input transition through multiple current-starved delay cells. If the delay of the $i$th cell is represented as $t_{d,i}$, the total accumulated delay is given by
\begin{equation}
    t_{d,\mathrm{acc}} = \sum_{i=1}^{N} t_{d,i},
\end{equation}
where $N$ is the number of multiplication cells being accumulated.

Ideally, each delay cell should exhibit a linear voltage-to-delay characteristic with respect to the multiplication output voltage $V_{\mathrm{MAC},i}$:
\begin{equation}
    t_{d,i} = t_0 + \alpha V_{\mathrm{MAC},i},
\end{equation}
where $t_0$ is the nominal delay and $\alpha$ is the voltage-to-delay gain. In this ideal case, the accumulated delay becomes
\begin{equation}
    t_{d,\mathrm{acc}} = N t_0 + \alpha \sum_{i=1}^{N} V_{\mathrm{MAC},i},
\end{equation}
which directly represents the accumulated MAC result in the time domain.

However, in practice, the current-starved delay cell has a nonlinear voltage-to-delay characteristic. This can be modeled as
\begin{equation}
    t_{d,i} = t_0 + \alpha V_{\mathrm{MAC},i}
    + \beta V_{\mathrm{MAC},i}^{2}
    + \gamma V_{\mathrm{MAC},i}^{3} + \cdots ,
\end{equation}
where $\beta$ and $\gamma$ represent the second- and third-order nonlinear delay coefficients, respectively. Therefore, the accumulated delay becomes
\begin{equation}
    t_{d,\mathrm{acc}} =
    N t_0
    + \alpha \sum_{i=1}^{N} V_{\mathrm{MAC},i}
    + \beta \sum_{i=1}^{N} V_{\mathrm{MAC},i}^{2}
    + \gamma \sum_{i=1}^{N} V_{\mathrm{MAC},i}^{3}
    + \cdots .
\end{equation}

The desired MAC information is contained in the linear term,
\begin{equation}
    t_{d,\mathrm{signal}} = \alpha \sum_{i=1}^{N} V_{\mathrm{MAC},i},
\end{equation}
while the higher-order terms introduce distortion:
\begin{equation}
    t_{d,\mathrm{dist}} =
    \beta \sum_{i=1}^{N} V_{\mathrm{MAC},i}^{2}
    + \gamma \sum_{i=1}^{N} V_{\mathrm{MAC},i}^{3}
    + \cdots .
\end{equation}

This shows that, in the cascaded delay-line architecture, delay-cell nonlinearities accumulate along with the desired signal. As the matrix dimension increases, the accumulated distortion also increases, which degrades the linearity of the final time-domain output. Nevertheless, the cascaded architecture has an important advantage: the entire accumulation is completed in a single propagation event. Therefore, the approximate computation time is
\begin{equation}
    T_{\mathrm{op,delay}} \approx t_{d,\mathrm{acc}},
\end{equation}
without requiring per-cell counter readout or sequential timing control. This enables higher operating frequency and lower control overhead.

In the counter-based architecture, the delay of each multiplication cell is measured independently and then accumulated digitally. The delay measured from each cell is quantized by the counter clock period $T_{\mathrm{clk}}$. The digitized output for the $i$th cell can be expressed as
\begin{equation}
    D_i = \left\lfloor \frac{t_{d,i}}{T_{\mathrm{clk}}} \right\rfloor .
\end{equation}

The final accumulated digital output is then
\begin{equation}
    D_{\mathrm{out}} = \sum_{i=1}^{N} D_i
    = \sum_{i=1}^{N}
    \left\lfloor \frac{t_{d,i}}{T_{\mathrm{clk}}} \right\rfloor .
\end{equation}

Since each cell is measured independently, the delay error of one stage does not propagate through the subsequent delay stages. This improves linearity compared with direct delay-line cascading. However, the counter-based architecture introduces timing quantization error. If the quantization error of each delay measurement is assumed to be uniformly distributed over one clock period, its variance is given by
\begin{equation}
    \sigma_{q,t}^{2} = \frac{T_{\mathrm{clk}}^{2}}{12}.
\end{equation}

For $N$ independently measured delay cells, the accumulated timing quantization noise becomes
\begin{equation}
    \sigma_{q,t,\mathrm{acc}}^{2}
    = N\frac{T_{\mathrm{clk}}^{2}}{12}.
\end{equation}

Thus, improving the resolution of the counter-based architecture requires reducing $T_{\mathrm{clk}}$, or equivalently increasing the counter clock frequency. However, this increases the power consumption of the counter and associated timing logic. The dynamic power of the digital control and counter circuitry can be approximated as
\begin{equation}
    P_{\mathrm{dig}} \approx \alpha_{\mathrm{sw}} C_{\mathrm{dig}} V_{\mathrm{DD}}^{2} f_{\mathrm{clk}},
\end{equation}
where $\alpha_{\mathrm{sw}}$ is the switching activity factor, $C_{\mathrm{dig}}$ is the effective switched capacitance, $V_{\mathrm{DD}}$ is the supply voltage, and $f_{\mathrm{clk}}$ is the counter clock frequency. Therefore, improving timing resolution directly increases digital power.

The counter-based architecture also requires sequential evaluation of the delay cells. If each cell requires a measurement time of approximately $T_{\mathrm{meas}}$, the total operation time becomes
\begin{equation}
    T_{\mathrm{op,counter}} \approx N T_{\mathrm{meas}} + T_{\mathrm{ctrl}},
\end{equation}
where $T_{\mathrm{ctrl}}$ represents the additional timing overhead from control logic, reset generation, and cell selection. This sequential operation reduces the maximum achievable throughput compared with the cascaded delay-line architecture.

Therefore, the cascaded delay-line architecture offers higher throughput and lower control overhead, but its accuracy is limited by accumulated delay nonlinearity. In contrast, the counter-based architecture improves linearity by independently digitizing each delay cell, but this comes at the cost of increased timing-control complexity, higher digital power, and lower operating frequency.

With additional circuit-level optimization, the cascaded delay-line architecture can be made more attractive for high-speed operation. For example, delay-cell linearization, calibration, bias optimization, or digital post-correction can be used to reduce the nonlinear distortion terms $\beta$ and $\gamma$ while preserving the high-throughput advantage of direct time-domain accumulation. Therefore, although the cascaded delay-line architecture exhibits stronger linearity limitations in its basic implementation, it can be preferable for high-frequency and energy-efficient operation when appropriate compensation techniques are included. Conversely, the counter-based architecture is more suitable when linearity and robustness are prioritized over speed and power efficiency.

\section{Simulation Results}
Fig.~9 and 10 compares the linearity of the cascaded delay-line accumulation architecture with the ideal response. The simulated delay outputs were extracted and post-processed in Python to obtain the measured output code. The result shows increasing deviation from the ideal linear response at higher MAC codes due to accumulated delay-cell nonlinearity. In contrast, Fig.~10 shows that the counter-based accumulation architecture closely follows the ideal response, with only quantization-limited staircase behavior introduced by the counter resolution.

\begin{figure}[htbp]
    \centering
    \includegraphics[scale=0.13]{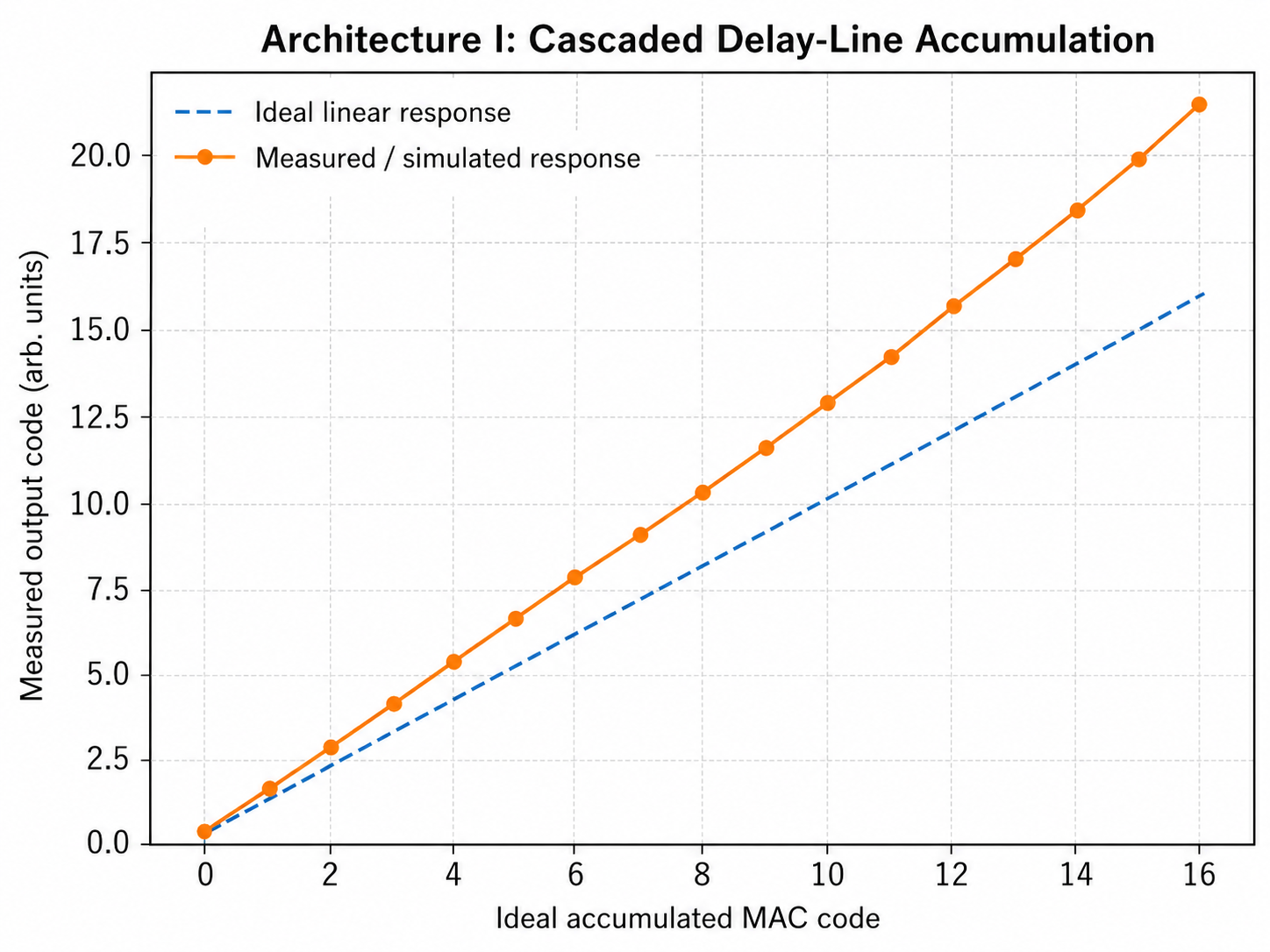} 
    \caption{Cascaded Delay Cell Based Macro}
    \label{fig:8}
\end{figure} 

\begin{figure}[htbp]
    \centering
    \includegraphics[scale=0.13]{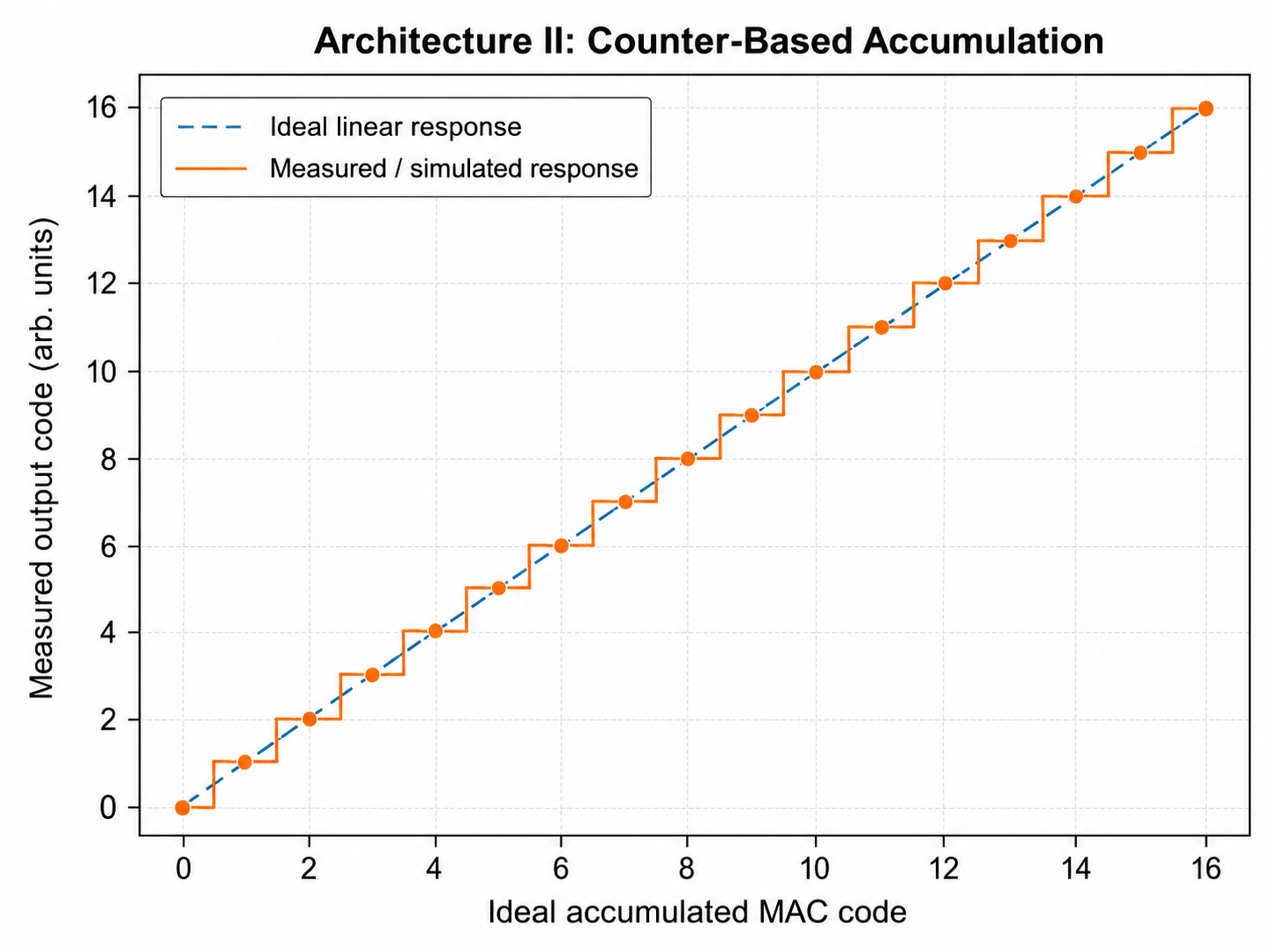} 
    \caption{Counter Based Macro}
    \label{fig:9}
\end{figure}

\begin{table*}[htbp]
\centering
\caption{Performance Comparison with Prior Work}
\label{tab:comparison}

\renewcommand{\arraystretch}{1.5} 

\begin{tabular*}{\textwidth}{@{\extracolsep{\fill}}|l|c|c|c|c|c|c|}
\hline
\textbf{} & \textbf{ISCAS'25[4]} & \textbf{ISCAS'22[10]} & \textbf{ISSCC'16[7]} & \textbf{ASSCC'16[8]} & \textbf{ISSC'20[9]} & \textbf{This Work (Macro 1)} \\
\hline
Process & 65 nm & 28 nm & 40 nm & 28 nm FDSOI & 28 nm & 65 nm \\
\hline
Supply & 1 V & 1 V & 1.1 V & 1 V & 1 V & 1 V \\
\hline
Clock rate & 300 MSps & 1 MHz & 2.5 GHz & 2.4 MHz & 50 MHz & 40 MHz \\
\hline
Input & Analog & 8 bit & 6 bit & 8 bit & 5 bit & 4 bit \\
\hline
Weights & 3x3 pixel (5b res) & 8 bit & 3 bit & - & 1 bit &  4 bit \\
\hline
Power & 120 uW & 29.3 uW & 331 uW & 7.74 uW & 74.79 uW &  42uW\\
\hline
Energy Efficiency & - &  1.04 TOPS/W &  7.7 TOPS/W & 9.61 TOPS/W & 1.34 TOPS/W  & 7.62 TOPS/W \\
\hline
Application & Edge Detection & NMC & Analog Accelerator & Image Classification & CNN & NMC \\
\hline
Feature & Time Domain & Time domain & Charge Domain & Charge Domain & Time Domain & Time Domain \\
\hline
Measured/Simulated & Measured & Measured & Measured & Measured & Simulated & Simulated \\
\hline
\end{tabular*}
\end{table*}

\section{Conclusion and Future Work}

This work presented two time-domain near-memory computing architectures for low-precision MAC operation. The first architecture performs accumulation using cascaded current-starved delay cells, while the second architecture uses a counter-based accumulation scheme. Although the counter-based architecture provides improved linearity by measuring each delay cell independently, it introduces significant power and timing overhead due to the additional control logic, counters, and sequential measurement operation. As a result, it operates at a lower effective frequency compared with the cascaded delay-line architecture.

In the present implementation, the achievable resolution remains limited to 4 bits, which restricts the applicability of the design for practical AI workloads requiring higher precision. Therefore, the performance numbers reported in this work are based on the first architecture, which provides higher operating frequency and lower control overhead despite its linearity limitations.

Future work will focus on improving the linearity and scalability of the delay-cell-based architecture. One approach is to regulate the PMOS current-starving device using an LDO-assisted biasing scheme, which can improve the linearity of the delay-control characteristic. In addition, RC-based delay cells will be investigated and compared against the current-starved inverter delay cell to evaluate their suitability for more linear voltage-to-time conversion. Further work will also target extending the architecture from the current 4-bit limitation toward 8-bit operation by improving the delay-cell linearity, optimizing the voltage-to-time conversion range, and developing calibration or compensation techniques for scalable time-domain accumulation.

\section*{Acknowledgment}

This project was completed to fulfill the course requirements of Advanced VLSI Design (CRN: ECE 69500 160). The authors would like to thank Prof. Sumeet Gupta for his guidance throughout the project and for his encouragement to explore the emerging field of near-memory computing.

\end{document}